\documentclass[aps,apl,twocolumn,groupedaddress,floatfix,letterpaper,superscriptaddress]{revtex4-2}
\usepackage{ae}
\usepackage[utf8]{inputenc}
\usepackage{graphicx}
\usepackage{hyperref}
\usepackage[leqno]{amsmath}
\usepackage{amsfonts}
\begin{document}
\title{Single-ion microwave near-field quantum sensor}

\author{M.~Wahnschaffe}
\affiliation{Physikalisch-Technische Bundesanstalt, Bundesallee 100, 38116 Braunschweig, Germany}
\affiliation{Institute of Quantum Optics, Leibniz Universität Hannover, Welfengarten 1, 30167 Hannover, Germany}
\affiliation{Laboratory for Nano- and Quantum Engineering, Leibniz Universität Hannover, Schneiderberg 39, 30167 Hannover, Germany}
\author{H.~Hahn}
\affiliation{Physikalisch-Technische Bundesanstalt, Bundesallee 100, 38116 Braunschweig, Germany}
\affiliation{Institute of Quantum Optics, Leibniz Universität Hannover, Welfengarten 1, 30167 Hannover, Germany}
\author{G.~Zarantonello}
\affiliation{Physikalisch-Technische Bundesanstalt, Bundesallee 100, 38116 Braunschweig, Germany}
\affiliation{Institute of Quantum Optics, Leibniz Universität Hannover, Welfengarten 1, 30167 Hannover, Germany}
\author{T.~Dubielzig}
\affiliation{Institute of Quantum Optics, Leibniz Universität Hannover, Welfengarten 1, 30167 Hannover, Germany}
\author{S.~Grondkowski}
\affiliation{Institute of Quantum Optics, Leibniz Universität Hannover, Welfengarten 1, 30167 Hannover, Germany}
\author{A.~Bautista-Salvador}
\affiliation{Physikalisch-Technische Bundesanstalt, Bundesallee 100, 38116 Braunschweig, Germany}
\affiliation{Institute of Quantum Optics, Leibniz Universität Hannover, Welfengarten 1, 30167 Hannover, Germany}
\affiliation{Laboratory for Nano- and Quantum Engineering, Leibniz Universität Hannover, Schneiderberg 39, 30167 Hannover, Germany}
\author{M.~Kohnen}
\affiliation{Physikalisch-Technische Bundesanstalt, Bundesallee 100, 38116 Braunschweig, Germany}
\affiliation{Institute of Quantum Optics, Leibniz Universität Hannover, Welfengarten 1, 30167 Hannover, Germany}
\affiliation{Laboratory for Nano- and Quantum Engineering, Leibniz Universität Hannover, Schneiderberg 39, 30167 Hannover, Germany}
\author{C.~Ospelkaus}
\affiliation{Physikalisch-Technische Bundesanstalt, Bundesallee 100, 38116 Braunschweig, Germany}
\affiliation{Institute of Quantum Optics, Leibniz Universität Hannover, Welfengarten 1, 30167 Hannover, Germany}
\affiliation{Laboratory for Nano- and Quantum Engineering, Leibniz Universität Hannover, Schneiderberg 39, 30167 Hannover, Germany}

\begin{abstract}
We develop an intuitive model of 2D microwave near-fields in the unusual regime of centimeter waves localized to tens of microns. Close to an intensity minimum, a simple effective description emerges with five parameters which characterize the strength and spatial orientation of the zero and first order terms of the near-field, as well as the field polarization. Such a field configuration is realized in a microfabricated planar structure with an integrated microwave conductor operating near 1\,GHz. We use a single $^9$Be$^+$ ion as a high-resolution quantum sensor to measure the field distribution through energy shifts in its hyperfine structure. We find agreement with simulations at the sub-micron and few-degree level. Our findings give a clear and general picture of the basic properties of oscillatory 2D near-fields with applications in quantum information processing, neutral atom trapping and manipulation, chip-scale atomic clocks, and integrated microwave circuits.
\end{abstract}

\maketitle

Static or oscillatory electromagnetic fields have important applications in atomic and molecular physics for atom trapping and manipulation. Neutral atoms can be trapped in static magnetic fields in different types of magnetic traps~\cite{ketterle_making_1999}. Atomic ions can be trapped either in superpositions of static and oscillatory electric fields (Paul trap) or in superimposed static electromagnetic fields (Penning trap)~\cite{ghosh_ion_1996}. Atom and molecule decelerators rely on the distortion of atomic energy levels by spatially inhomogeneous fields~\cite{van_de_meerakker_manipulation_2012}. Common to all of these field configurations is that their basic properties can be well described in terms of static solutions to the field equations and that the behavior of the field near its intensity minimum is often critical to the application. Prominent examples include Majorana losses in neutral atom magnetic traps~\cite{ketterle_making_1999} and micromotion in Paul traps~\cite{berkeland_minimization_1998}. 

Recently, motivated by advances in microfabricated atom traps, interest has grown in microwave near-fields which originate from  microfabricated structures. Dimensions are typically small compared to the wavelength, but for the relatively high frequencies involved, eddy currents and phase effects become important, and the resulting field patterns are much richer than in the quasistatic case. Examples include rf potentials for neutral atoms~\cite{fortagh_magnetic_2007} with applications in atom interferometry, quantum gates~\cite{calarco_quantum_2000,treutlein_microwave_2006} and chip-scale atomic clocks~\cite{maineult_spin_2012} as well as microwave near-fields for trapped-ion quantum logic~\cite{ospelkaus_trapped-ion_2008,ospelkaus_microwave_2011,allcock_microfabricated_2013}. Also, neutral atomic clouds~\cite{bohi_imaging_2010,horsley_widefield_2015} and single ions~\cite{warring_techniques_2013} have been used to characterize near-fields at sub-mm length scales, measure magnetic field gradients~\cite{walther_single_2011} or for microwave magnetometry~\cite{baumgart_ultrasensitive_2016}. 
The behavior of these high-frequency oscillatory fields may also become relevant for coupling atomic and molecular quantum systems to microwave circuits in the quantum regime~\cite{schuster_cavity_2011,kielpinski_quantum_2012}. Of particular importance in this context are 2D field configurations which can be realized e.~g.~in integrated waveguides. Notwithstanding the strong experimental interest, there is a lack of intuitive understanding and the wide-spread notion that numerical simulation of microwave near-fields originating from such structures is difficult due to the many inductive and capacitive couplings between conductors. 

Here we develop a simple picture of 2D microwave fields around a local minimum of the field intensity and confirm this model through numerical simulations and experimental measurements involving a microfabricated ion trap with an integrated microwave conductor. We assume that the dimensions are small compared to the wavelength, so that $\mathrm{div} \vec B=0$ and $\mathrm{rot} \vec B=0$ (near-field condition). Expansion of a 2D field up to first order would in principle result in a total of 6 complex or 12 real-valued expansion coefficients. However, taking into account the near-field condition, we can write the magnetic field in terms of eight parameters: $B_{r,i}$ and $\alpha_{r,i}$, characterizing the real and imaginary components of the complex field at the origin and their spatial orientations, and $B'_{r,i}$ and $\beta_{r,i}$, which describe the real and imaginary components of the complex field gradient and their spatial orientations:
\begin{multline}
	\label{eq:2dquadrupsimp}
	\vec B = \mathrm{Re}
	         \bigg\{
	           e^{i\omega t}
	           \Big[
	             \left( B_r  \vec e_{\alpha_r} + i B_i \vec e_{\alpha_i} \right) + \\
	             \left( B'_r      Q_{\beta_r}  + i B'_i     Q_{\beta_i}  \right)\vec r + \ldots
               \Big]
             \bigg\},
	         \\
	\vec e_{\alpha}\equiv
	\begin{pmatrix}
		\cos\alpha \\
		\sin\alpha
	\end{pmatrix}
	\quad\mathrm{and}\quad
	Q_\beta\equiv
	\begin{pmatrix}
		\cos\beta &  \sin\beta \\
		\sin\beta & -\cos\beta \\
	\end{pmatrix},
\end{multline}
where $Q_\beta$ is a traceless and symmetric ``quadrupole matrix'' to ensure the near-field condition. By multiplying Eq.~(\ref{eq:2dquadrupsimp}) with a suitably chosen complex phase factor, it is possible to maximize the strength of the real part of the gradient. The same choice of phase factor also leads to $\beta_i=\beta_r-\pi/2$. We now write $(B_r,B_i)=B(\cos\varphi,\sin\varphi)$ and $(B'_r,B'_i)\equiv B'(\cos\psi,\sin\psi)$. A suitable choice for the domain of the parameters is $B,\,B'\in\mathbb{R}$, $\alpha_r,\beta_r,\psi \in \left[0,\pi\right[$, $\alpha_i,\beta_i,\varphi \in \left[-\pi/2,\pi/2\right[$. Further imposing the condition that $|\vec B|$ has a minimum at the origin leads to $\alpha_i-\alpha_r+\pi/2=n\cdot\pi$ with $n \in \mathbb{Z}$. For our choice of parameters, the left-hand side must be in $\left]-\pi,\pi\right[$, and thus $n=0$ and also $\alpha_i=\alpha_r-\pi/2$. Also from $|\vec B|$ minimal at the origin, we find $\varphi=\psi-\pi/2$. With $\alpha\equiv\alpha_r$ and $\beta\equiv\beta_r$, the field is finally given by 
\begin{multline}
  \label{eq:2dquadrup}
  \vec B = 
  \mathrm{Re}
  \bigg\{
	e^{i\omega t}
	\Big[
      B \left( \vec e_{\alpha}\sin\psi - i \vec e_{\alpha-\pi/2}\cos\psi \right) + \\
      B'\left(      Q_{\beta} \cos\psi + i      Q_{\beta-\pi/2} \sin\psi \right)\vec r +\ldots
    \Big]
  \bigg\}
\end{multline}
with just five free parameters -- the strengths $B$ and $B'$ of the offset field and of the gradient, respectively, one angle $\alpha$ and $\beta$ each for their spatial orientation, and an angle $\psi$ characterizing the relative strength of the real and imaginary part of the gradient (and thus the polarization). The reduction from eight to five parameters compared to Eq.~(\ref{eq:2dquadrupsimp}) is due to the assumption of a specific phase and of a minimum of $|\vec B|$ at the origin.

\begin{figure}[tb]
	\centering
	\includegraphics[width=\columnwidth]{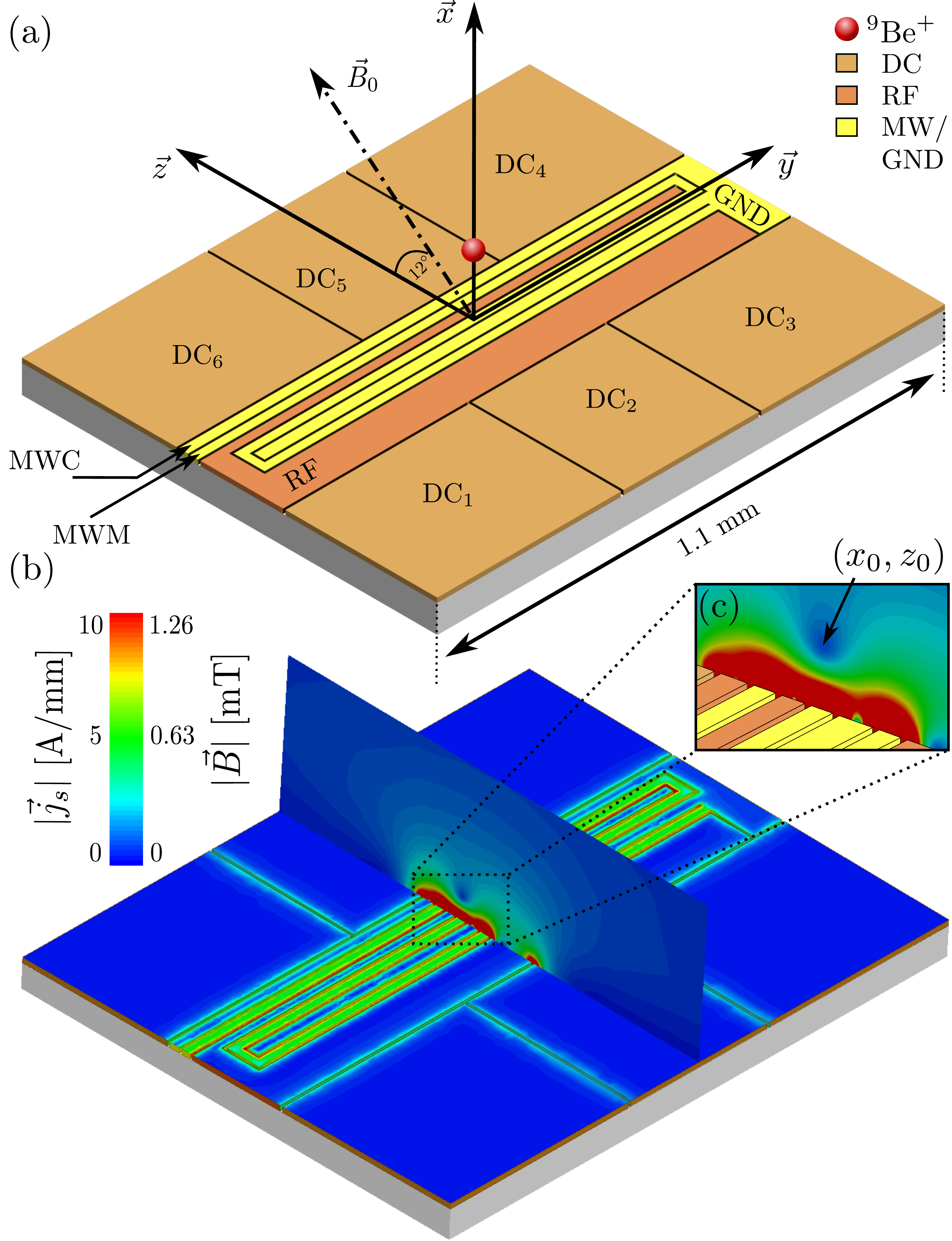}
	\caption{(a) Surface-electrode ion trap structure. DC and RF voltages applied to the bright and dark orange electrodes create a harmonic trapping potential for a single ion at the position indicated by the red sphere. A microwave current coupled into the conductor MWM (yellow) leads to the surface-current distribution $|\vec j_\mathrm{s}|$ depicted in (b). The resulting magnetic near-field $|\vec B|$ is shown in the $xz$ plane (close-up in (c)). Around $(x_0,z_0)$, the near-field is described by the model of Eq.~(\ref{eq:2dquadrup}) and characterized using a single ion as a microwave quantum sensor. For clarity, the height of the ion above the surface has been exaggerated in (a).}
	\label{fig:figure1}
\end{figure}

To give a specific example, consider the surface-electrode trap structure shown in Fig.~\ref{fig:figure1}(a), a design evolved from~\cite{carsjens_surface-electrode_2014}. It is located in a room temperature vacuum enclosure evacuated to $\approx 1\cdot 10^{-11}\,\mathrm{mbar}$. The trap is composed of $11\,\mu\mathrm{m}$ thick electroplated gold electrodes (yellow, bright and dark orange) with insulating $5\,\mu\mathrm{m}$ wide gaps (black lines) between the electrodes on top of an insulating AlN substrate (gray)~\cite{treutlein_coherent_2008}. A single $^9$Be$^+$ ion is trapped above the surface by DC and RF electric fields. These are generated by applying a radio-frequency voltage ($2\pi\cdot 88$\,MHz, $100\,\mathrm{V_{pp}}$) to the electrode (RF), resulting in ponderomotive forces pushing the ion towards $(x_p,z_p)=(45.7,2.9)\,\mu\mathrm{m}$. Additional DC voltages applied to electrodes $\mathrm{DC}_{1-6}$ push the ion towards $y=0$, but may also create additional forces in the $xz$ plane. The latter let us fine-tune the position of the ion, indicated by the red sphere, in the $xz$ plane. The trap depth is 39\,meV, and the trap frequencies are given by $\omega_y\simeq2\pi\cdot 1$\,MHz and $\omega_{x,z}\simeq2\pi\cdot 11$\,MHz. 

In addition, microwave conductors, shown in yellow in Fig.~\ref{fig:figure1}(a), are integrated into the structure for quantum state control of trapped ions. For this purpose, it is desirable to achieve a near-field pattern as described by Eq.~(\ref{eq:2dquadrup}) with $B/B'$ as small as possible at a position where the ion can be trapped. These near-fields can then be used to implement multi-qubit quantum logic gates for quantum information processing with trapped ions~\cite{ospelkaus_trapped-ion_2008,ospelkaus_microwave_2011}. Towards this end, we apply a microwave current at 1.093\,GHz to the conductor MWM. Fig.~\ref{fig:figure1}(b) shows the corresponding simulated surface current distribution $|\vec j_\mathrm{s}|$ in the electrode structure. A slice in the $xz$ plane shows the resulting magnetic near-field $|\vec B|$ for $y=0$. Fig.~\ref{fig:figure1}(c) shows a close-up of the distribution of $|\vec B|$ around $(x_0,z_0)\approx(45.5,-0.9)\,\mu\mathrm{m}$, where it exhibits a local minimum. Here we show that around $(x_0,z_0)$, this near-field is accurately described by Eq.~(\ref{eq:2dquadrup}). We characterize the field distribution using a single ion as a quantum sensor and show agreement with numerical simulations of $\vec B$. 

We simulate the structure, including parts of the surrounding connector board, using Ansys HFSS. The simulations deliver $\vec B$ on a grid in the $xz$ plane. The simulations show that $B_y$ is much smaller than $B_x$ and $B_z$, which validates the assumption of a 2D field configuration. We thus fit the model of Eq.~(\ref{eq:2dquadrup}) to the numerical $B_x$ and $B_z$ data on a $3\,\mu\mathrm{m}$ by $3\,\mu\mathrm{m}$ square to extract the parameters of Eq.~(\ref{eq:2dquadrup}). Here, in Eq.~(\ref{eq:2dquadrup}), we substitute $\vec r$ by $\vec r - (x_0,z_0)^T$, as the local field minimum is not located at the origin, and obtain the values of $x_0,z_0$ as additional fit parameters. The resulting parameters are shown in Table \ref{tab:table1}. Note that $B$ and $B'$ depend on the input current, and hence only $B/B'$ is given. Our simulations show a rather small value for $\psi$; as a result, the real part of the quadrupole is much stronger than the imaginary part. Hence, the polarization is mostly linear. The dominant contribution to the gradient $B'$ stems from the three conductor segments forming the meander MWM, while the offset field $B$, which is $\pi/2$ out of phase with the gradient, results from inductive coupling to neighboring metal electrodes and from the associated eddy currents visible in Fig.~\ref{fig:figure1}(b), as well as from phase delays along the meander \cite{carsjens_surface-electrode_2014}. The errors of the fit parameters are the standard errors from the nonlinear least squares fit and indicate how well the model of Eq.~(\ref{eq:2dquadrup}) describes the field distribution. 

\begin{figure}[tb]
	\centering
	\includegraphics[width=0.8\columnwidth]{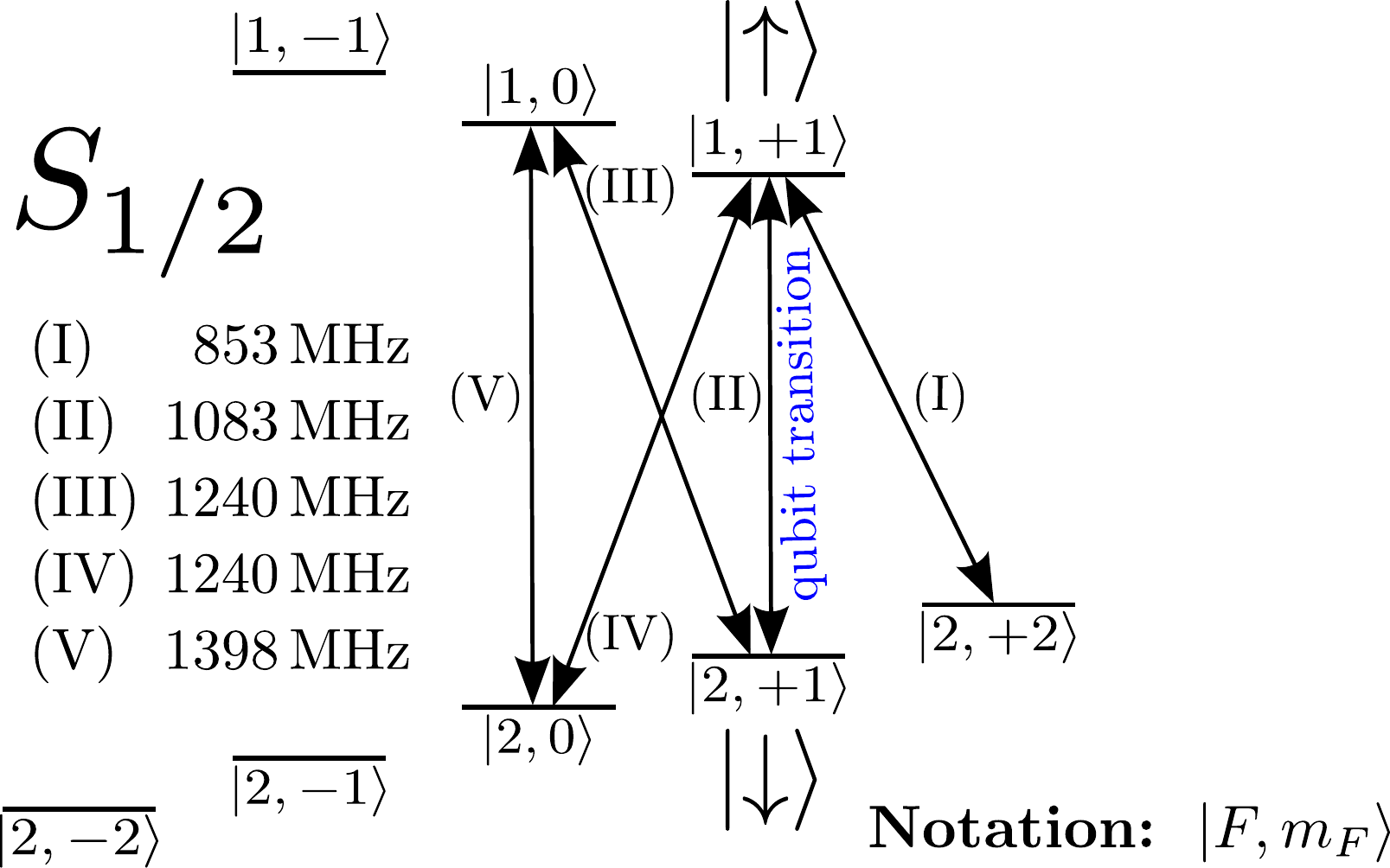}
	\caption{Hyperfine structure of the $^9$Be$^+$ ground state at 22.3\,mT, where transition (II) is a first-order magnetic-field independent qubit transition. }
	\label{fig:figure2}
\end{figure}

The magnetic near-field $\vec B$ primarily results in energy shifts of internal hyperfine states of the $^9$Be$^+$ ion and does not affect its position significantly. The main idea of the experiment is to measure these shifts spectroscopically for different positions of the ion controlled by the DC voltages. We can thus determine the parameters of Eq.~(\ref{eq:2dquadrup}) experimentally and compare them to the simulations. We load single ions into the trap by hitting a solid $^9$Be target with single pulses of a nanosecond pulsed laser at 1064\,nm and by subsequent resonant two-photon ionization at 235\,nm~\cite{lo_all-solid-state_2013,hendricks_all-optical_2007} from the resulting ablation plume. Ions are laser cooled and detected using light resonant with the cycling transition $\left|S_{1/2},F=2,\,m_F=2\right>\rightarrow\left|P_{3/2},m_J=+3/2,m_I=+3/2\right>$ at 313\,nm. We apply a static bias field $\vec B_0$ in the $yz$ plane and at an angle of $12^\circ$ with respect to the $z$ axis to lift the degeneracy of the hyperfine levels. The hyperfine sublevels of the ground state are shown in Fig.~\ref{fig:figure2} and labeled with $\left|F,\,m_F\right>$. Here, $F$ is the quantum number of the total angular momentum $\vec F$ and $m_F$ the quantum number of its projection on $\vec B_0$. At the experimental value of $B_0=22.3\,\mathrm{mT}$, the state combination $\left|F=2,\,m_F=1\right>$ and $\left|F=1,\,m_F=1\right>$ forms a first order magnetic-field independent qubit~\cite{langer_long-lived_2005} which can be exploited for long coherence times. Laser cooling prepares the ion in $\left|F=2,\,m_F=2\right>$. Through a series of microwave current pulses on the conductor MWC (cf.~Fig.\ref{fig:figure1}(a)), resonant with suitable hyperfine transitions, we can prepare an arbitrary target state within the $S_{1/2}$ hyperfine manifold of Fig.~\ref{fig:figure2}, and determine the population of any state by transferring it back to $\left|F=2,\,m_F=2\right>$ and subsequently detecting fluorescence photons scattered on the cycling transition. 

We determine properties of the microwave near-field through AC Zeeman energy shifts which it induces on suitable atomic hyperfine states, analogous to AC Stark shifts for optical fields. The AC Zeeman shift of a hyperfine energy level $E_i$ is given by 
\begin{equation}
  \delta E_\mathrm{AC}=\hbar\cdot\mathrm{sgn}(E_j-E_i)\sum_{j\ne i} \frac{|\Omega_{ij}(\vec B)|^2}{\omega-\omega_{ij}},
\end{equation}
where the sum is over all other energy levels $j$, $\omega_{ij}=|E_i-E_j|/\hbar$ and $\Omega_{ij}(\vec B)$ is the Rabi rate for the $i\leftrightarrow j$ transition which depends linearly on atomic matrix elements and the components of $\vec B$. For actual calculations, we also take into account the (small) Bloch-Siegert shift from the counter-rotating term. Plugging in $\vec B$ from Eq.~(\ref{eq:2dquadrup}), the AC Zeeman shift of a hyperfine energy level is a polynomial in $x-x_0$ and $z-z_0$ (up to second order in each), where the coefficients are built from atomic matrix elements, trigonometric functions of $\psi,\alpha,\beta$, and $B$ and $B'$. We perform a nonlinear least-squares fit of this expression to experimentally measured AC Zeeman shifts as a function of $x,z$ to obtain $\psi,\alpha,\beta,B,B',x_0$ and $z_0$. Experimentally, we cannot measure absolute energies, only relative shifts of two energy levels by probing the transition frequency between them. The shift of the transition frequency thus has the same form as the AC Zeeman shift of an individual level, just more terms. We denote these as $\delta f_{\mathrm{AC},k}(x,z;B,B',\psi,\alpha,\beta,x_0,z_0)$, where $k$ identifies a transition in Fig.~\ref{fig:figure2}, for example $k=\mathrm{(II)}$.

We simultaneously fit datasets for two different transitions within the structure of Fig.~\ref{fig:figure2}, (V) and (II), because they couple to the polarization components of the field differently and thus provide complementary information. The (II) data exhibits a strong sensitivity to $\alpha,\beta,\psi$, whereas the (V) data is mainly sensitive to $B'$, $B$, $x_0$ and $z_0$. We first test this procedure on numerical HFSS data from which we calculate the expected AC Zeeman shifts on a useful grid of ion positions ($x,z$). We simultaneously fit $\delta f_\mathrm{AC,(II)}$ and $\delta f_\mathrm{AC,(V)}$ to this simulated AC Zeeman shift data. We find perfect agreement between the field parameters obtained from the simulated AC Zeeman shift data and those extracted directly from the fit of Eq.~(\ref{eq:2dquadrup}) to the simulation data (Table \ref{tab:table1}). 

\begin{table}[tb]
	\begin{ruledtabular}
		\begin{tabular}{lll}
			Parameter [units]  & Simulation                  & Experimental data\\
			$B/B' \,[\mu\mathrm{m}]$    & $8.20(2)$    & $8.7(1.0)$\\
			$\psi \,[^\circ]$     & $6.5(1)$              & $4.3(1.2)$\\
			$\alpha \,[^\circ]$   & $25.15(2)$            & $31.1(3)$\\	
			$\beta \,[^\circ]$    & $99.3(1)$             & $109.1(11.5)$\\
			$x_0  \,[\mu\mathrm{m}]$      & $45.46(2)$   & $45.3(1)$\\
			$z_0  \,[\mu\mathrm{m}]$      & $-0.855(6)$  & $-0.8(2)$
		\end{tabular}
	\end{ruledtabular}
	\caption{Parameters of the microwave near-fields according to Eq.~(\ref{eq:2dquadrup}), determined from simulations and from experimental measurements using a single  $^9$Be$^+$ ion (Fig.~\ref{fig:figure3}).}
	\label{tab:table1} 
\end{table}

\begin{figure}[tb]
	\centering
	\includegraphics[width=\columnwidth]{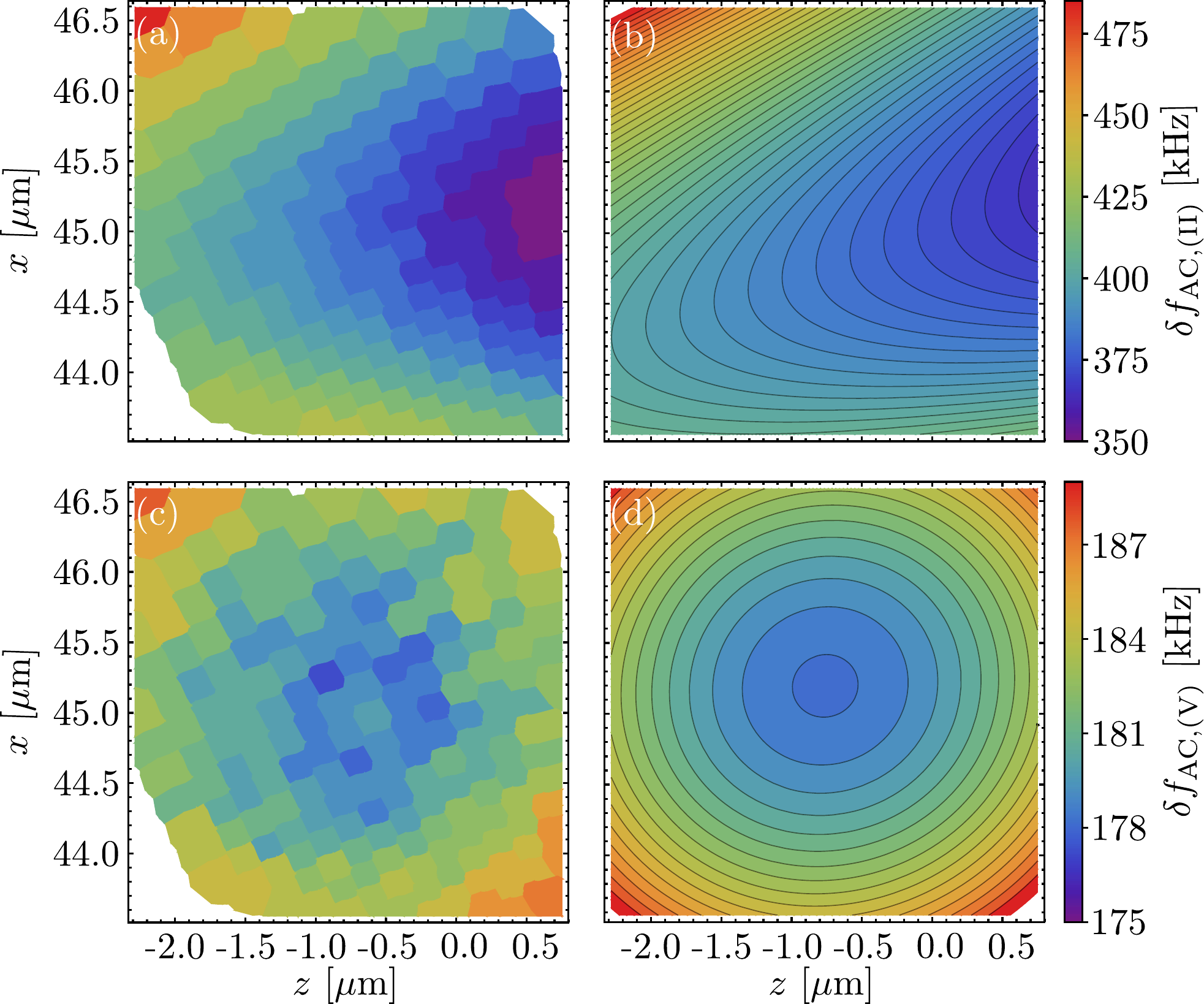}
	\caption{AC Zeeman shifts $\delta f_{\mathrm{AC},k}$ induced by 2D near-fields on a single ion. The two rows shows data for transition (II) and (V) of Fig.~\ref{fig:figure2}, respectively. The first column shows AC Zeeman shifts measured using a single ion, and the second column shows result of a fit of Eq.~(\ref{eq:2dquadrup}) to the experimental data. Empty (white) areas in the radial plane indicate where we cannot stably trap ions.}
	\label{fig:figure3}
\end{figure}

While in principle such shifts could be measured using Rabi spectroscopy, we employ the Ramsey method described in~\cite{warring_techniques_2013} because it lends itself to easy automation. It does, however, not directly reveal the sign of the Zeeman shifts. In the following, we will therefore always show positive signs of the net shifts. The first column of Fig.~\ref{fig:figure3} shows AC Zeeman shifts of transitions (II) (top) and (V) (bottom) as a function of $x$ and $z$, measured using a single ion. For transition (V), the AC Zeeman shift should exhibit a minimum close to the minimum of $|\vec B|$. As can be seen from Fig.~\ref{fig:figure3}(a), the data for transition (II) exhibits a more complex structure, which is a result of terms with different signs adding up in the total AC Zeeman shift calculation. We fit $\delta f_\mathrm{AC,(II)}$ and $\delta f_\mathrm{AC,(V)}$ to this data to obtain the fit parameters given in the third column of Table \ref{tab:table1}. The calculated AC Zeeman shifts resulting from the fitted model are plotted in the right column of Fig.~\ref{fig:figure3}. Data for transition (V) was taken at a power level that was nominally 6\,dB higher than for (II) in order to reach higher frequency shifts. Thus, we also fitted the experimental power ratio between Fig.~\ref{fig:figure3}(c) and (a), yielding 6.47(15)\,dB. Experimental and fitted data have been scaled to the power level of Fig.~\ref{fig:figure3}(a). For reference, Fig.~\ref{fig:figure3}(a) and (b) correspond to $B'\approx45\,\mathrm{T/m}$, while the data for Fig.~\ref{fig:figure3}(c) and (d) was taken for $B'\approx94\,\mathrm{T/m}$ and was then scaled to match Fig.~\ref{fig:figure3}(a) and (b) as described above. As can be seen from Table \ref{tab:table1}, the agreement between simulations and experiment is at the sub-micron and few-degree level. This is remarkable given the complicated interplay of primary and induced currents in this microfabricated structure where the properties of the field around the minimum essentially result from the subtraction of rather large contributions from individual conductors~\cite{ospelkaus_trapped-ion_2008,carsjens_surface-electrode_2014}. 

An issue which may cause the fitted parameters to deviate slightly from the simulations is the accuracy of the assumed spatial position of the ion as a function of trap voltages applied. The position was extracted from electrostatic simulations and the pseudopotential approximation. Also, in our simulations, we found that spurious couplings to the electrode MWC had a rather strong influence on $B/B'$ and on $(x_0, z_0)$. There is a $\approx10\%$ coupling from the MWM to the MWC conductor. The value of $B/B'$ therefore depends on the assumed termination on the MWC input. For our simulations, we assumed that about 5\% of the total power coupled from the MWM to the MWC conductor is reflected back into the structure. This is not an unreasonably high value, given a number of impedance changes which occur between the structure of Fig.~\ref{fig:figure1} and the amplifier connected to MWC. Additional frequency shifts as a result of a potentially inhomogeneous bias field $B_0$ or spurious oscillatory magnetic fields associated with the RF trap drive might arise. However, these should be fully canceled by the spin-echo sequence employed in the spectroscopy~\cite{warring_techniques_2013}. 

In summary, we have developed an intuitive model of 2D microwave quadrupole fields around a local minimum of $|\vec B|$, performed accurate numerical simulations of a 2D near-field structure, and confirmed their accuracy at the sub-micron and few-degree level using a single ion as a local field probe. The field model of Eq.~(\ref{eq:2dquadrup}) is essential, as it allows us to compare simulations with experimental data. This description is applicable not only to microwave, but also to lower frequency rf fields. Our results will inform the design of advanced structures for microwave quantum logic applications~\cite{mintert_ion-trap_2001,ospelkaus_trapped-ion_2008,ospelkaus_microwave_2011,khromova_designer_2012,allcock_microfabricated_2013,lake_generation_2015} of trapped ions. The model delivers a figure of merit, $B/B'$, and parameters ($\alpha,\beta,\psi$), directly relevant for this application. Ideally, future designs would be based on a multi-layer structure~\cite{amini_toward_2010,guise_ball-grid_2015,moehring_design_2011}, so that signals could be delivered in separated layers underneath the structure via embedded waveguides and only brought to the surface close to the ion~\cite{craik_microwave_2014}. This would decouple the design of near-field structures from other trap ``modules'' on a scalable trap array~\cite{mielenz_freely_2015} for quantum simulation~\cite{blatt_quantum_2012,schneider_experimental_2012} or quantum logic applications~\cite{wineland_nobel_2013,monroe_scaling_2013}. One can also interpret our measurements as a nanometer range resolution quantum enabled microwave magnetic field probe. The methodology developed here, combined with a ``stylus'' ion trap~\cite{maiwald_stylus_2009}, could be used to characterize micro antennas and waveguides. Our findings may be applicable to integrated microwave circuits and hybrid quantum approaches coupling ions to other microwave or rf quantum devices~\cite{schuster_cavity_2011,kielpinski_quantum_2012}. 

\begin{acknowledgments}
We are grateful for useful discussions with P.~O.~Schmidt, the NIST ion storage group, the Oxford ion trapping group and J.~Sch\"obel. We acknowledge contributions by C.~Vollmer in an early stage of the experiment, support by the PTB clean room facility team, packaging support by F.~Schmidt-Kaler's group, and funding from QUEST, PTB, LUH, NTH (project number 2.2.11) and DFG through CRC 1227 DQ-\textit{mat}.   
\end{acknowledgments} 


\begin{thebibliography}{36}
	\expandafter\ifx\csname natexlab\endcsname\relax\def\natexlab#1{#1}\fi
	\expandafter\ifx\csname bibnamefont\endcsname\relax
	\def\bibnamefont#1{#1}\fi
	\expandafter\ifx\csname bibfnamefont\endcsname\relax
	\def\bibfnamefont#1{#1}\fi
	\expandafter\ifx\csname citenamefont\endcsname\relax
	\def\citenamefont#1{#1}\fi
	\expandafter\ifx\csname url\endcsname\relax
	\def\url#1{\texttt{#1}}\fi
	\expandafter\ifx\csname urlprefix\endcsname\relax\def\urlprefix{URL }\fi
	\providecommand{\bibinfo}[2]{#2}
	\providecommand{\eprint}[2][]{\url{#2}}
	
	\bibitem[{\citenamefont{Ketterle et~al.}(1999)\citenamefont{Ketterle, Durfee,
			and Stamper-Kurn}}]{ketterle_making_1999}
	\bibinfo{author}{\bibfnamefont{W.}~\bibnamefont{Ketterle}},
	\bibinfo{author}{\bibfnamefont{D.~S.} \bibnamefont{Durfee}},
	\bibnamefont{and} \bibinfo{author}{\bibfnamefont{D.~M.}
		\bibnamefont{Stamper-Kurn}}, in \emph{\bibinfo{booktitle}{Proceedings of the
			1998 {Enrico} {Fermi} summer school on {Bose}-{Einstein} condensation in
			{Varenna}, {Italy}}} (\bibinfo{publisher}{IOS Press},
	\bibinfo{address}{Amsterdam}, \bibinfo{year}{1999}), pp.
	\bibinfo{pages}{67--176}.
	
	\bibitem[{\citenamefont{Ghosh}(1996)}]{ghosh_ion_1996}
	\bibinfo{author}{\bibfnamefont{P.~K.} \bibnamefont{Ghosh}},
	\emph{\bibinfo{title}{Ion {Traps}}}, International {Series} of {Monographs}
	on {Physics} (\bibinfo{publisher}{Clarendon Press},
	\bibinfo{address}{Oxford}, \bibinfo{year}{1996}).
	
	\bibitem[{\citenamefont{van~de Meerakker et~al.}(2012)\citenamefont{van~de
			Meerakker, Bethlem, Vanhaecke, and
			Meijer}}]{van_de_meerakker_manipulation_2012}
	\bibinfo{author}{\bibfnamefont{S.~Y.~T.} \bibnamefont{van~de Meerakker}},
	\bibinfo{author}{\bibfnamefont{H.~L.} \bibnamefont{Bethlem}},
	\bibinfo{author}{\bibfnamefont{N.}~\bibnamefont{Vanhaecke}},
	\bibnamefont{and} \bibinfo{author}{\bibfnamefont{G.}~\bibnamefont{Meijer}},
	\bibinfo{journal}{Chemical Reviews} \textbf{\bibinfo{volume}{112}},
	\bibinfo{pages}{4828} (\bibinfo{year}{2012}).
	
	\bibitem[{\citenamefont{Berkeland et~al.}(1998)\citenamefont{Berkeland, Miller,
			Bergquist, Itano, and Wineland}}]{berkeland_minimization_1998}
	\bibinfo{author}{\bibfnamefont{D.~J.} \bibnamefont{Berkeland}},
	\bibinfo{author}{\bibfnamefont{J.~D.} \bibnamefont{Miller}},
	\bibinfo{author}{\bibfnamefont{J.~C.} \bibnamefont{Bergquist}},
	\bibinfo{author}{\bibfnamefont{W.~M.} \bibnamefont{Itano}}, \bibnamefont{and}
	\bibinfo{author}{\bibfnamefont{D.~J.} \bibnamefont{Wineland}},
	\bibinfo{journal}{Journal of Applied Physics} \textbf{\bibinfo{volume}{83}},
	\bibinfo{pages}{5025} (\bibinfo{year}{1998}).
	
	\bibitem[{\citenamefont{Fortágh and Zimmermann}(2007)}]{fortagh_magnetic_2007}
	\bibinfo{author}{\bibfnamefont{J.}~\bibnamefont{Fortágh}} \bibnamefont{and}
	\bibinfo{author}{\bibfnamefont{C.}~\bibnamefont{Zimmermann}},
	\bibinfo{journal}{Reviews of Modern Physics} \textbf{\bibinfo{volume}{79}},
	\bibinfo{pages}{235} (\bibinfo{year}{2007}).
	
	\bibitem[{\citenamefont{Calarco et~al.}(2000)\citenamefont{Calarco, Hinds,
			Jaksch, Schmiedmayer, Cirac, and Zoller}}]{calarco_quantum_2000}
	\bibinfo{author}{\bibfnamefont{T.}~\bibnamefont{Calarco}},
	\bibinfo{author}{\bibfnamefont{E.~A.} \bibnamefont{Hinds}},
	\bibinfo{author}{\bibfnamefont{D.}~\bibnamefont{Jaksch}},
	\bibinfo{author}{\bibfnamefont{J.}~\bibnamefont{Schmiedmayer}},
	\bibinfo{author}{\bibfnamefont{J.~I.} \bibnamefont{Cirac}}, \bibnamefont{and}
	\bibinfo{author}{\bibfnamefont{P.}~\bibnamefont{Zoller}},
	\bibinfo{journal}{Physical Review A} \textbf{\bibinfo{volume}{61}},
	\bibinfo{pages}{022304} (\bibinfo{year}{2000}).
	
	\bibitem[{\citenamefont{Treutlein et~al.}(2006)\citenamefont{Treutlein,
			Hänsch, Reichel, Negretti, Cirone, and Calarco}}]{treutlein_microwave_2006}
	\bibinfo{author}{\bibfnamefont{P.}~\bibnamefont{Treutlein}},
	\bibinfo{author}{\bibfnamefont{T.~W.} \bibnamefont{Hänsch}},
	\bibinfo{author}{\bibfnamefont{J.}~\bibnamefont{Reichel}},
	\bibinfo{author}{\bibfnamefont{A.}~\bibnamefont{Negretti}},
	\bibinfo{author}{\bibfnamefont{M.~A.} \bibnamefont{Cirone}},
	\bibnamefont{and} \bibinfo{author}{\bibfnamefont{T.}~\bibnamefont{Calarco}},
	\bibinfo{journal}{Physical Review A} \textbf{\bibinfo{volume}{74}},
	\bibinfo{pages}{022312} (\bibinfo{year}{2006}).
	
	\bibitem[{\citenamefont{Maineult et~al.}(2012)\citenamefont{Maineult, Deutsch,
			Gibble, Reichel, and Rosenbusch}}]{maineult_spin_2012}
	\bibinfo{author}{\bibfnamefont{W.}~\bibnamefont{Maineult}},
	\bibinfo{author}{\bibfnamefont{C.}~\bibnamefont{Deutsch}},
	\bibinfo{author}{\bibfnamefont{K.}~\bibnamefont{Gibble}},
	\bibinfo{author}{\bibfnamefont{J.}~\bibnamefont{Reichel}}, \bibnamefont{and}
	\bibinfo{author}{\bibfnamefont{P.}~\bibnamefont{Rosenbusch}},
	\bibinfo{journal}{Physical Review Letters} \textbf{\bibinfo{volume}{109}},
	\bibinfo{pages}{020407} (\bibinfo{year}{2012}).
	
	\bibitem[{\citenamefont{Ospelkaus et~al.}(2008)\citenamefont{Ospelkaus, Langer,
			Amini, Brown, Leibfried, and Wineland}}]{ospelkaus_trapped-ion_2008}
	\bibinfo{author}{\bibfnamefont{C.}~\bibnamefont{Ospelkaus}},
	\bibinfo{author}{\bibfnamefont{C.~E.} \bibnamefont{Langer}},
	\bibinfo{author}{\bibfnamefont{J.~M.} \bibnamefont{Amini}},
	\bibinfo{author}{\bibfnamefont{K.~R.} \bibnamefont{Brown}},
	\bibinfo{author}{\bibfnamefont{D.}~\bibnamefont{Leibfried}},
	\bibnamefont{and} \bibinfo{author}{\bibfnamefont{D.~J.}
		\bibnamefont{Wineland}}, \bibinfo{journal}{Physical Review Letters}
	\textbf{\bibinfo{volume}{101}}, \bibinfo{pages}{090502}
	(\bibinfo{year}{2008}).
	
	\bibitem[{\citenamefont{Ospelkaus et~al.}(2011)\citenamefont{Ospelkaus,
			Warring, Colombe, Brown, Amini, Leibfried, and
			Wineland}}]{ospelkaus_microwave_2011}
	\bibinfo{author}{\bibfnamefont{C.}~\bibnamefont{Ospelkaus}},
	\bibinfo{author}{\bibfnamefont{U.}~\bibnamefont{Warring}},
	\bibinfo{author}{\bibfnamefont{Y.}~\bibnamefont{Colombe}},
	\bibinfo{author}{\bibfnamefont{K.~R.} \bibnamefont{Brown}},
	\bibinfo{author}{\bibfnamefont{J.~M.} \bibnamefont{Amini}},
	\bibinfo{author}{\bibfnamefont{D.}~\bibnamefont{Leibfried}},
	\bibnamefont{and} \bibinfo{author}{\bibfnamefont{D.~J.}
		\bibnamefont{Wineland}}, \bibinfo{journal}{Nature}
	\textbf{\bibinfo{volume}{476}}, \bibinfo{pages}{181} (\bibinfo{year}{2011}).
	
	\bibitem[{\citenamefont{Allcock et~al.}(2013)\citenamefont{Allcock, Harty,
			Ballance, Keitch, Linke, Stacey, and Lucas}}]{allcock_microfabricated_2013}
	\bibinfo{author}{\bibfnamefont{D.~T.~C.} \bibnamefont{Allcock}},
	\bibinfo{author}{\bibfnamefont{T.~P.} \bibnamefont{Harty}},
	\bibinfo{author}{\bibfnamefont{C.~J.} \bibnamefont{Ballance}},
	\bibinfo{author}{\bibfnamefont{B.~C.} \bibnamefont{Keitch}},
	\bibinfo{author}{\bibfnamefont{N.~M.} \bibnamefont{Linke}},
	\bibinfo{author}{\bibfnamefont{D.~N.} \bibnamefont{Stacey}},
	\bibnamefont{and} \bibinfo{author}{\bibfnamefont{D.~M.} \bibnamefont{Lucas}},
	\bibinfo{journal}{Applied Physics Letters} \textbf{\bibinfo{volume}{102}},
	\bibinfo{pages}{044103} (\bibinfo{year}{2013}).
	
	\bibitem[{\citenamefont{Böhi et~al.}(2010)\citenamefont{Böhi, Riedel,
			Hänsch, and Treutlein}}]{bohi_imaging_2010}
	\bibinfo{author}{\bibfnamefont{P.}~\bibnamefont{Böhi}},
	\bibinfo{author}{\bibfnamefont{M.~F.} \bibnamefont{Riedel}},
	\bibinfo{author}{\bibfnamefont{T.~W.} \bibnamefont{Hänsch}},
	\bibnamefont{and}
	\bibinfo{author}{\bibfnamefont{P.}~\bibnamefont{Treutlein}},
	\bibinfo{journal}{Applied Physics Letters} \textbf{\bibinfo{volume}{97}},
	\bibinfo{pages}{051101} (\bibinfo{year}{2010}).
	
	\bibitem[{\citenamefont{Horsley et~al.}(2015)\citenamefont{Horsley, Du, and
			Treutlein}}]{horsley_widefield_2015}
	\bibinfo{author}{\bibfnamefont{A.}~\bibnamefont{Horsley}},
	\bibinfo{author}{\bibfnamefont{G.-X.} \bibnamefont{Du}}, \bibnamefont{and}
	\bibinfo{author}{\bibfnamefont{P.}~\bibnamefont{Treutlein}},
	\bibinfo{journal}{New Journal of Physics} \textbf{\bibinfo{volume}{17}},
	\bibinfo{pages}{112002} (\bibinfo{year}{2015}).
	
	\bibitem[{\citenamefont{Warring et~al.}(2013)\citenamefont{Warring, Ospelkaus,
			Colombe, Brown, Amini, Carsjens, Leibfried, and
			Wineland}}]{warring_techniques_2013}
	\bibinfo{author}{\bibfnamefont{U.}~\bibnamefont{Warring}},
	\bibinfo{author}{\bibfnamefont{C.}~\bibnamefont{Ospelkaus}},
	\bibinfo{author}{\bibfnamefont{Y.}~\bibnamefont{Colombe}},
	\bibinfo{author}{\bibfnamefont{K.~R.} \bibnamefont{Brown}},
	\bibinfo{author}{\bibfnamefont{J.~M.} \bibnamefont{Amini}},
	\bibinfo{author}{\bibfnamefont{M.}~\bibnamefont{Carsjens}},
	\bibinfo{author}{\bibfnamefont{D.}~\bibnamefont{Leibfried}},
	\bibnamefont{and} \bibinfo{author}{\bibfnamefont{D.~J.}
		\bibnamefont{Wineland}}, \bibinfo{journal}{Physical Review A}
	\textbf{\bibinfo{volume}{87}}, \bibinfo{pages}{013437}
	(\bibinfo{year}{2013}).
	
	\bibitem[{\citenamefont{Walther et~al.}(2011)\citenamefont{Walther, Poschinger,
			Ziesel, Hettrich, Wiens, Welzel, and Schmidt-Kaler}}]{walther_single_2011}
	\bibinfo{author}{\bibfnamefont{A.}~\bibnamefont{Walther}},
	\bibinfo{author}{\bibfnamefont{U.}~\bibnamefont{Poschinger}},
	\bibinfo{author}{\bibfnamefont{F.}~\bibnamefont{Ziesel}},
	\bibinfo{author}{\bibfnamefont{M.}~\bibnamefont{Hettrich}},
	\bibinfo{author}{\bibfnamefont{A.}~\bibnamefont{Wiens}},
	\bibinfo{author}{\bibfnamefont{J.}~\bibnamefont{Welzel}}, \bibnamefont{and}
	\bibinfo{author}{\bibfnamefont{F.}~\bibnamefont{Schmidt-Kaler}},
	\bibinfo{journal}{Physical Review A} \textbf{\bibinfo{volume}{83}},
	\bibinfo{pages}{062329} (\bibinfo{year}{2011}).
	
	\bibitem[{\citenamefont{Baumgart et~al.}(2016)\citenamefont{Baumgart, Cai,
			Retzker, Plenio, and Wunderlich}}]{baumgart_ultrasensitive_2016}
	\bibinfo{author}{\bibfnamefont{I.}~\bibnamefont{Baumgart}},
	\bibinfo{author}{\bibfnamefont{J.-M.} \bibnamefont{Cai}},
	\bibinfo{author}{\bibfnamefont{A.}~\bibnamefont{Retzker}},
	\bibinfo{author}{\bibfnamefont{M.}~\bibnamefont{Plenio}}, \bibnamefont{and}
	\bibinfo{author}{\bibfnamefont{C.}~\bibnamefont{Wunderlich}},
	\bibinfo{journal}{Physical Review Letters} \textbf{\bibinfo{volume}{116}},
	\bibinfo{pages}{240801} (\bibinfo{year}{2016}).
	
	\bibitem[{\citenamefont{Schuster et~al.}(2011)\citenamefont{Schuster, Bishop,
			Chuang, DeMille, and Schoelkopf}}]{schuster_cavity_2011}
	\bibinfo{author}{\bibfnamefont{D.~I.} \bibnamefont{Schuster}},
	\bibinfo{author}{\bibfnamefont{L.~S.} \bibnamefont{Bishop}},
	\bibinfo{author}{\bibfnamefont{I.~L.} \bibnamefont{Chuang}},
	\bibinfo{author}{\bibfnamefont{D.}~\bibnamefont{DeMille}}, \bibnamefont{and}
	\bibinfo{author}{\bibfnamefont{R.~J.} \bibnamefont{Schoelkopf}},
	\bibinfo{journal}{Physical Review A} \textbf{\bibinfo{volume}{83}},
	\bibinfo{pages}{012311} (\bibinfo{year}{2011}).
	
	\bibitem[{\citenamefont{Kielpinski et~al.}(2012)\citenamefont{Kielpinski,
			Kafri, Woolley, Milburn, and Taylor}}]{kielpinski_quantum_2012}
	\bibinfo{author}{\bibfnamefont{D.}~\bibnamefont{Kielpinski}},
	\bibinfo{author}{\bibfnamefont{D.}~\bibnamefont{Kafri}},
	\bibinfo{author}{\bibfnamefont{M.~J.} \bibnamefont{Woolley}},
	\bibinfo{author}{\bibfnamefont{G.~J.} \bibnamefont{Milburn}},
	\bibnamefont{and} \bibinfo{author}{\bibfnamefont{J.~M.}
		\bibnamefont{Taylor}}, \bibinfo{journal}{Physical Review Letters}
	\textbf{\bibinfo{volume}{108}}, \bibinfo{pages}{130504}
	(\bibinfo{year}{2012}).
	
	\bibitem[{\citenamefont{Carsjens et~al.}(2014)\citenamefont{Carsjens, Kohnen,
			Dubielzig, and Ospelkaus}}]{carsjens_surface-electrode_2014}
	\bibinfo{author}{\bibfnamefont{M.}~\bibnamefont{Carsjens}},
	\bibinfo{author}{\bibfnamefont{M.}~\bibnamefont{Kohnen}},
	\bibinfo{author}{\bibfnamefont{T.}~\bibnamefont{Dubielzig}},
	\bibnamefont{and}
	\bibinfo{author}{\bibfnamefont{C.}~\bibnamefont{Ospelkaus}},
	\bibinfo{journal}{Applied Physics B} \textbf{\bibinfo{volume}{114}},
	\bibinfo{pages}{243} (\bibinfo{year}{2014}).
	
	\bibitem[{\citenamefont{Treutlein}(2008)}]{treutlein_coherent_2008}
	\bibinfo{author}{\bibfnamefont{P.}~\bibnamefont{Treutlein}}, Ph.D. thesis,
	\bibinfo{school}{Ludwig-Maximilians-Universität München}
	(\bibinfo{year}{2008}).
	
	\bibitem[{\citenamefont{Lo et~al.}(2013)\citenamefont{Lo, Alonso, Kienzler,
			Keitch, Clercq, Negnevitsky, and Home}}]{lo_all-solid-state_2013}
	\bibinfo{author}{\bibfnamefont{H.-Y.} \bibnamefont{Lo}},
	\bibinfo{author}{\bibfnamefont{J.}~\bibnamefont{Alonso}},
	\bibinfo{author}{\bibfnamefont{D.}~\bibnamefont{Kienzler}},
	\bibinfo{author}{\bibfnamefont{B.~C.} \bibnamefont{Keitch}},
	\bibinfo{author}{\bibfnamefont{L.~E.~d.} \bibnamefont{Clercq}},
	\bibinfo{author}{\bibfnamefont{V.}~\bibnamefont{Negnevitsky}},
	\bibnamefont{and} \bibinfo{author}{\bibfnamefont{J.~P.} \bibnamefont{Home}},
	\bibinfo{journal}{Applied Physics B} \textbf{\bibinfo{volume}{114}},
	\bibinfo{pages}{17} (\bibinfo{year}{2013}).
	
	\bibitem[{\citenamefont{Hendricks et~al.}(2007)\citenamefont{Hendricks, Grant,
			Herskind, Dantan, and Drewsen}}]{hendricks_all-optical_2007}
	\bibinfo{author}{\bibfnamefont{R.~J.} \bibnamefont{Hendricks}},
	\bibinfo{author}{\bibfnamefont{D.~M.} \bibnamefont{Grant}},
	\bibinfo{author}{\bibfnamefont{P.~F.} \bibnamefont{Herskind}},
	\bibinfo{author}{\bibfnamefont{A.}~\bibnamefont{Dantan}}, \bibnamefont{and}
	\bibinfo{author}{\bibfnamefont{M.}~\bibnamefont{Drewsen}},
	\bibinfo{journal}{Applied Physics B} \textbf{\bibinfo{volume}{88}},
	\bibinfo{pages}{507} (\bibinfo{year}{2007}).
	
	\bibitem[{\citenamefont{Langer et~al.}(2005)\citenamefont{Langer, Ozeri, Jost,
			Chiaverini, DeMarco, Ben-Kish, Blakestad, Britton, Hume, Itano
			et~al.}}]{langer_long-lived_2005}
	\bibinfo{author}{\bibfnamefont{C.}~\bibnamefont{Langer}},
	\bibinfo{author}{\bibfnamefont{R.}~\bibnamefont{Ozeri}},
	\bibinfo{author}{\bibfnamefont{J.~D.} \bibnamefont{Jost}},
	\bibinfo{author}{\bibfnamefont{J.}~\bibnamefont{Chiaverini}},
	\bibinfo{author}{\bibfnamefont{B.}~\bibnamefont{DeMarco}},
	\bibinfo{author}{\bibfnamefont{A.}~\bibnamefont{Ben-Kish}},
	\bibinfo{author}{\bibfnamefont{R.~B.} \bibnamefont{Blakestad}},
	\bibinfo{author}{\bibfnamefont{J.}~\bibnamefont{Britton}},
	\bibinfo{author}{\bibfnamefont{D.~B.} \bibnamefont{Hume}},
	\bibinfo{author}{\bibfnamefont{W.~M.} \bibnamefont{Itano}},
	\bibnamefont{et~al.}, \bibinfo{journal}{Physical Review Letters}
	\textbf{\bibinfo{volume}{95}}, \bibinfo{pages}{060502}
	(\bibinfo{year}{2005}).
	
	\bibitem[{\citenamefont{Mintert and Wunderlich}(2001)}]{mintert_ion-trap_2001}
	\bibinfo{author}{\bibfnamefont{F.}~\bibnamefont{Mintert}} \bibnamefont{and}
	\bibinfo{author}{\bibfnamefont{C.}~\bibnamefont{Wunderlich}},
	\bibinfo{journal}{Physical Review Letters} \textbf{\bibinfo{volume}{87}},
	\bibinfo{pages}{257904} (\bibinfo{year}{2001}).
	
	\bibitem[{\citenamefont{Khromova et~al.}(2012)\citenamefont{Khromova, Piltz,
			Scharfenberger, Gloger, Johanning, Varón, and
			Wunderlich}}]{khromova_designer_2012}
	\bibinfo{author}{\bibfnamefont{A.}~\bibnamefont{Khromova}},
	\bibinfo{author}{\bibfnamefont{C.}~\bibnamefont{Piltz}},
	\bibinfo{author}{\bibfnamefont{B.}~\bibnamefont{Scharfenberger}},
	\bibinfo{author}{\bibfnamefont{T.~F.} \bibnamefont{Gloger}},
	\bibinfo{author}{\bibfnamefont{M.}~\bibnamefont{Johanning}},
	\bibinfo{author}{\bibfnamefont{A.~F.} \bibnamefont{Varón}},
	\bibnamefont{and}
	\bibinfo{author}{\bibfnamefont{C.}~\bibnamefont{Wunderlich}},
	\bibinfo{journal}{Physical Review Letters} \textbf{\bibinfo{volume}{108}},
	\bibinfo{pages}{220502} (\bibinfo{year}{2012}).
	
	\bibitem[{\citenamefont{Lake et~al.}(2015)\citenamefont{Lake, Weidt, Randall,
			Standing, Webster, and Hensinger}}]{lake_generation_2015}
	\bibinfo{author}{\bibfnamefont{K.}~\bibnamefont{Lake}},
	\bibinfo{author}{\bibfnamefont{S.}~\bibnamefont{Weidt}},
	\bibinfo{author}{\bibfnamefont{J.}~\bibnamefont{Randall}},
	\bibinfo{author}{\bibfnamefont{E.~D.} \bibnamefont{Standing}},
	\bibinfo{author}{\bibfnamefont{S.~C.} \bibnamefont{Webster}},
	\bibnamefont{and} \bibinfo{author}{\bibfnamefont{W.~K.}
		\bibnamefont{Hensinger}}, \bibinfo{journal}{Physical Review A}
	\textbf{\bibinfo{volume}{91}}, \bibinfo{pages}{012319}
	(\bibinfo{year}{2015}).
	
	\bibitem[{\citenamefont{Amini et~al.}(2010)\citenamefont{Amini, Uys, Wesenberg,
			Seidelin, Britton, Bollinger, Leibfried, Ospelkaus, VanDevender, and
			Wineland}}]{amini_toward_2010}
	\bibinfo{author}{\bibfnamefont{J.~M.} \bibnamefont{Amini}},
	\bibinfo{author}{\bibfnamefont{H.}~\bibnamefont{Uys}},
	\bibinfo{author}{\bibfnamefont{J.~H.} \bibnamefont{Wesenberg}},
	\bibinfo{author}{\bibfnamefont{S.}~\bibnamefont{Seidelin}},
	\bibinfo{author}{\bibfnamefont{J.}~\bibnamefont{Britton}},
	\bibinfo{author}{\bibfnamefont{J.~J.} \bibnamefont{Bollinger}},
	\bibinfo{author}{\bibfnamefont{D.}~\bibnamefont{Leibfried}},
	\bibinfo{author}{\bibfnamefont{C.}~\bibnamefont{Ospelkaus}},
	\bibinfo{author}{\bibfnamefont{A.~P.} \bibnamefont{VanDevender}},
	\bibnamefont{and} \bibinfo{author}{\bibfnamefont{D.~J.}
		\bibnamefont{Wineland}}, \bibinfo{journal}{New Journal of Physics}
	\textbf{\bibinfo{volume}{12}}, \bibinfo{pages}{033031}
	(\bibinfo{year}{2010}).
	
	\bibitem[{\citenamefont{Guise et~al.}(2015)\citenamefont{Guise, Fallek,
			Stevens, Brown, Volin, Harter, Amini, Higashi, Lu, Chanhvongsak
			et~al.}}]{guise_ball-grid_2015}
	\bibinfo{author}{\bibfnamefont{N.~D.} \bibnamefont{Guise}},
	\bibinfo{author}{\bibfnamefont{S.~D.} \bibnamefont{Fallek}},
	\bibinfo{author}{\bibfnamefont{K.~E.} \bibnamefont{Stevens}},
	\bibinfo{author}{\bibfnamefont{K.~R.} \bibnamefont{Brown}},
	\bibinfo{author}{\bibfnamefont{C.}~\bibnamefont{Volin}},
	\bibinfo{author}{\bibfnamefont{A.~W.} \bibnamefont{Harter}},
	\bibinfo{author}{\bibfnamefont{J.~M.} \bibnamefont{Amini}},
	\bibinfo{author}{\bibfnamefont{R.~E.} \bibnamefont{Higashi}},
	\bibinfo{author}{\bibfnamefont{S.~T.} \bibnamefont{Lu}},
	\bibinfo{author}{\bibfnamefont{H.~M.} \bibnamefont{Chanhvongsak}},
	\bibnamefont{et~al.}, \bibinfo{journal}{Journal of Applied Physics}
	\textbf{\bibinfo{volume}{117}}, \bibinfo{pages}{174901}
	(\bibinfo{year}{2015}).
	
	\bibitem[{\citenamefont{Moehring et~al.}(2011)\citenamefont{Moehring,
			Highstrete, Stick, Fortier, Haltli, Tigges, and
			Blain}}]{moehring_design_2011}
	\bibinfo{author}{\bibfnamefont{D.~L.} \bibnamefont{Moehring}},
	\bibinfo{author}{\bibfnamefont{C.}~\bibnamefont{Highstrete}},
	\bibinfo{author}{\bibfnamefont{D.}~\bibnamefont{Stick}},
	\bibinfo{author}{\bibfnamefont{K.~M.} \bibnamefont{Fortier}},
	\bibinfo{author}{\bibfnamefont{R.}~\bibnamefont{Haltli}},
	\bibinfo{author}{\bibfnamefont{C.}~\bibnamefont{Tigges}}, \bibnamefont{and}
	\bibinfo{author}{\bibfnamefont{M.~G.} \bibnamefont{Blain}},
	\bibinfo{journal}{New Journal of Physics} \textbf{\bibinfo{volume}{13}},
	\bibinfo{pages}{075018} (\bibinfo{year}{2011}).
	
	\bibitem[{\citenamefont{Craik et~al.}(2014)\citenamefont{Craik, Linke, Harty,
			Ballance, Lucas, Steane, and Allcock}}]{craik_microwave_2014}
	\bibinfo{author}{\bibfnamefont{D.~P. L.~A.} \bibnamefont{Craik}},
	\bibinfo{author}{\bibfnamefont{N.~M.} \bibnamefont{Linke}},
	\bibinfo{author}{\bibfnamefont{T.~P.} \bibnamefont{Harty}},
	\bibinfo{author}{\bibfnamefont{C.~J.} \bibnamefont{Ballance}},
	\bibinfo{author}{\bibfnamefont{D.~M.} \bibnamefont{Lucas}},
	\bibinfo{author}{\bibfnamefont{A.~M.} \bibnamefont{Steane}},
	\bibnamefont{and} \bibinfo{author}{\bibfnamefont{D.~T.~C.}
		\bibnamefont{Allcock}}, \bibinfo{journal}{Applied Physics B}
	\textbf{\bibinfo{volume}{114}}, \bibinfo{pages}{3} (\bibinfo{year}{2014}).
	
	\bibitem[{\citenamefont{Mielenz et~al.}(2015)\citenamefont{Mielenz, Kalis,
			Wittemer, Hakelberg, Schmied, Blain, Maunz, Leibfried, Warring, and
			Schaetz}}]{mielenz_freely_2015}
	\bibinfo{author}{\bibfnamefont{M.}~\bibnamefont{Mielenz}},
	\bibinfo{author}{\bibfnamefont{H.}~\bibnamefont{Kalis}},
	\bibinfo{author}{\bibfnamefont{M.}~\bibnamefont{Wittemer}},
	\bibinfo{author}{\bibfnamefont{F.}~\bibnamefont{Hakelberg}},
	\bibinfo{author}{\bibfnamefont{R.}~\bibnamefont{Schmied}},
	\bibinfo{author}{\bibfnamefont{M.}~\bibnamefont{Blain}},
	\bibinfo{author}{\bibfnamefont{P.}~\bibnamefont{Maunz}},
	\bibinfo{author}{\bibfnamefont{D.}~\bibnamefont{Leibfried}},
	\bibinfo{author}{\bibfnamefont{U.}~\bibnamefont{Warring}}, \bibnamefont{and}
	\bibinfo{author}{\bibfnamefont{T.}~\bibnamefont{Schaetz}},
	\bibinfo{journal}{arXiv:1512.03559 [physics, physics:quant-ph]}
	(\bibinfo{year}{2015}), \bibinfo{note}{arXiv: 1512.03559}.
	
	\bibitem[{\citenamefont{Blatt and Roos}(2012)}]{blatt_quantum_2012}
	\bibinfo{author}{\bibfnamefont{R.}~\bibnamefont{Blatt}} \bibnamefont{and}
	\bibinfo{author}{\bibfnamefont{C.~F.} \bibnamefont{Roos}},
	\bibinfo{journal}{Nature Physics} \textbf{\bibinfo{volume}{8}},
	\bibinfo{pages}{277} (\bibinfo{year}{2012}).
	
	\bibitem[{\citenamefont{Schneider et~al.}(2012)\citenamefont{Schneider, Porras,
			and Schaetz}}]{schneider_experimental_2012}
	\bibinfo{author}{\bibfnamefont{C.}~\bibnamefont{Schneider}},
	\bibinfo{author}{\bibfnamefont{D.}~\bibnamefont{Porras}}, \bibnamefont{and}
	\bibinfo{author}{\bibfnamefont{T.}~\bibnamefont{Schaetz}},
	\bibinfo{journal}{Reports on Progress in Physics}
	\textbf{\bibinfo{volume}{75}}, \bibinfo{pages}{024401}
	(\bibinfo{year}{2012}).
	
	\bibitem[{\citenamefont{Wineland}(2013)}]{wineland_nobel_2013}
	\bibinfo{author}{\bibfnamefont{D.~J.} \bibnamefont{Wineland}},
	\bibinfo{journal}{Reviews of Modern Physics} \textbf{\bibinfo{volume}{85}},
	\bibinfo{pages}{1103} (\bibinfo{year}{2013}).
	
	\bibitem[{\citenamefont{Monroe and Kim}(2013)}]{monroe_scaling_2013}
	\bibinfo{author}{\bibfnamefont{C.}~\bibnamefont{Monroe}} \bibnamefont{and}
	\bibinfo{author}{\bibfnamefont{J.}~\bibnamefont{Kim}},
	\bibinfo{journal}{Science} \textbf{\bibinfo{volume}{339}},
	\bibinfo{pages}{1164} (\bibinfo{year}{2013}).
	
	\bibitem[{\citenamefont{Maiwald et~al.}(2009)\citenamefont{Maiwald, Leibfried,
			Britton, Bergquist, Leuchs, and Wineland}}]{maiwald_stylus_2009}
	\bibinfo{author}{\bibfnamefont{R.}~\bibnamefont{Maiwald}},
	\bibinfo{author}{\bibfnamefont{D.}~\bibnamefont{Leibfried}},
	\bibinfo{author}{\bibfnamefont{J.}~\bibnamefont{Britton}},
	\bibinfo{author}{\bibfnamefont{J.~C.} \bibnamefont{Bergquist}},
	\bibinfo{author}{\bibfnamefont{G.}~\bibnamefont{Leuchs}}, \bibnamefont{and}
	\bibinfo{author}{\bibfnamefont{D.~J.} \bibnamefont{Wineland}},
	\bibinfo{journal}{Nature Physics} \textbf{\bibinfo{volume}{5}},
	\bibinfo{pages}{551} (\bibinfo{year}{2009}).
	
\end{thebibliography}

\end{document}